\documentclass[aps]{revtex4}
\usepackage{color}
\usepackage{graphicx}
\usepackage{eurosym}
\usepackage{graphicx}
\usepackage{amsmath,amssymb}
\usepackage{color}
\usepackage[outdir=./]{epstopdf}
\usepackage{amsmath}
\usepackage{amsfonts}
\usepackage{amssymb}

\def\bc{\begin{center}}
\def\ec{\end{center}}

\def\beq{\begin{eqnarray}}
\def\eeq{\end{eqnarray}}

\def\bc{\begin{center}}
\def\ec{\end{center}}

\def\beq{\begin{eqnarray}}
\def\eeq{\end{eqnarray}}

\input{tcilatex}
\begin{document}

\title{Inflation deployed torus-shaped solar sail accelerated via thermal
desorption of coating}
\date{\today }
\author{Roman Ya. Kezerashvili$^{1,2}$, Olga L. Starinova$^{3}$, Alexander
S. Chekashov$^{3}$, Dylan J. Slocki$^{4}$}
\affiliation{\mbox{$^{1}$New York
City College
of Technology, The City University of New York,} \\
Brooklyn, NY 11201, USA \\
\mbox{$^{2}$The Graduate School and University Center, The
City University of New York,} \\
New York, NY 10016, USA \\
\mbox{$^{3}$Samara National Research University, Russian Federation, Samara,
Russia }\\
\mbox{$^{4}$State University of New York at Buffalo, Buffalo, NY 14228, USA}}
\date{\today}

\begin{abstract}
A torus-shaped sail consists of a reflective membrane attached to an
inflatable torus-shaped rim. The sail's deployment from its stowed
configuration is initiated by introducing inflation pressure into the
toroidal rim with an attached circular flat membrane coated by
heat-sensitive materials that undergo thermal desorption (TD) from a solid
to a gas phase. Our study of the deployment and acceleration of the sail is
split into three steps: at a particular heliocentric distance a torus-shaped
sail is deployed by a gas inflated into the toroidal rim and the membrane is
kept flat by the pressure of the gas; under heating by solar radiation, the
membrane coat undergoes TD and the sail is accelerated via TD of coating and
solar radiation pressure (SRP); when TD ends, the sail utilizes thrust only
from SRP. We study the stability of the torus-shaped sail and deflection and
vibration of the flat membrane due to the acceleration by TD and SRP. The
stability of the toroidal rim is addressed.
\end{abstract}

\maketitle


\section{Introduction}

Electromagnetic radiation of the Sun is the abundant source of energy in
space, it can provide spacecraft with a gentle yet persistent thrust for
interplanetary and interstellar missions by means of a solar sail. Solar
sail spacecraft offer an alternative to the traditional propulsion systems
that must carry their fuel on the vehicle. A solar sail is a large sheet of
low areal density material that captures or reflects the Sun electromagnetic
flux as a means of acceleration \cite{pol, Colin, Matloff3, MatloffValpetti}%
. The solar sail navigates by changing the orientation or reflectivity of
the sails. These factors benefit performance since the spacecraft no longer
needs to carry the mass of a propulsion system. This in turn benefits the
acceleration of the solar sail. The initial launch of a solar sailing
spacecraft is typically more affordable due to having a lighter weight. It
would also not require refueling missions that can limit the longevity of
the spacecraft. The proposals to use solar sails covers almost the whole
spectrum of space missions, from the construction of a lunar base to an
expedition to Mars, from scientific probes to mining exploitation of the
asteroids, even implementations for deep space exploration and interstellar
travel are considered.

Solar sails come in various types, typically being rigid flat squares with
mechanical booms that keep the sail material in place. A few of these beams
are under compressive loads with stay wires that keep the sail relatively
flat. Canopy sails are similar in construction to parachute sails or pillow
sails. They use the solar radiation pressure to remain open much like a
parachute uses air. Canopy sails decrease the required mass and stiffness in
relation to a decrease of performance. Less of the actual area is used for
propulsion as a result of increased curvature. Many different systems have
been previously considered for the sails opening. Each system was
characterized by the presence of guide rollers, electromechanical actuation
devices, or composite booms \cite{9}. Most recently an alternative method
for the solar sail self-deployment based on shape memory alloys was
suggested \cite{MemorySail, MemorySail2}, where the authors use of shape
memory alloys as mechanical actuators for solar sail self-deployment instead
of heavy and bulky mechanical booms. However, in the actual deployment
technology, the main limit is still the high weight of the system and the
complexity of the deployment mechanism for solar sail surface.

\bigskip The concept of using inflatable structures for spacecraft has been
extensively discussed during the past six decades and dates back to the
1950's. The Echo balloons launched in the 1960s by NASA were some of the
first inflatable to be utilized \cite{5}. This was followed by the
Inflatable Antenna Experiment around the middle of the 1990s \cite{6}.
Throughout the 1960s and early 1970s, NASA and industry teams were at work
developing inflatable space structures ranging from space suits to habitats
\cite{4}. Space inflatable composite structures offer great advantages over
rigid structures due to being flexible and more compatible low volume
storage for launch. Once these structures are inflated and deployed, they
operated with similar principals to rigid structures to obtain comparable or
greater performance. Inflatable structures are becoming more robust and
durable following the advancement and development of flexible polymers and
high strength fibers. This allows for increasingly smaller mass packed into
a dense prelaunch storage, which further reduces costs \cite{4}. Inflatable
space structures have a simple deployment mechanism for large space
structures besides being beneficial from their light-weight and small volume
for launch. Inflatable structures, which have been the subject of renewed
interest in recent years and investigated in detail in Refs. \cite{Leigh,
Tinker}, have characteristics that are particularly advantageous for a solar
sail. First, they are extremely lightweight, which makes them an ideal match
for use with low thrust solar sails, where sail weight is critical. The
second obvious advantage is the ability to deploy in orbit and related space
savings in the launch configuration. Therefore, inflatable structures
possess special properties such as low weight, minimal stowage volume and
high stretch-to-fold ratios that make them suitable for solar sails. In 1989
Jorg Strobl in Ref. \cite{Strobl1} proposed a hydrogen inflated hollow
disk-shaped sail with a molybdenum reflector. This hollow body solar sail
concept was later applied to the design of an orbiting radio telescope \cite%
{Strobl2}. Matloff \cite{Matloff1} reinvestigated and revived the
hollow-body solar sail an interstellar travel concept. Space-environment
effects on beryllium hollow-body sails with hydrogen fill gas unfurled was
investigated in details in Refs. \cite{Kezmetloff1,Kezmetloff2}. In Ref.\
\cite{Genta1999} the concept of the solar sail with inflatable beams was
suggested, which are kept pressurized after deployment, relieves all
compressive stresses, allowing a very simple configuration and a
straightforward deployment procedure. The solar sail with inflatable booms,
that unfold to support the sail and control vanes at the corners was
analyzed in Ref. \cite{8}. The idea of the inflatable ring sail or the torus
solar sail was presented in Ref. \cite{Hayn}, where a circular sail with an
inflatable outer ring was considered.

The acceleration of the solar sail by thermal desorption of a coating was
proposed by Benford and Benford \cite{Benford}. In this case the reflective
area of the sail should be coated by the material that undergoes the thermal
desorption when beam powered microwave pulse from the source located on the
Earth or on an orbit is used for heating. The experimental study \cite%
{Benford2} demonstrated that desorption can attain high specific impulse if
low mass molecules or atoms are blown out of a lattice of material at high
temperature. However, the solar sail is naturally heated through the
absorption of solar radiation. It was suggested to utilize the concept of
thermal desorption to the solar sail that naturally gains temperature
through the absorption of solar radiation at a particular point in a
heliocentric orbit where the temperature of the solar sail corresponds to
the temperature of thermal desorption of the coating material \cite{1}. It
is of particular interest to consider an inflatable torus-shaped solar sail
as both propellant-less and propellant-based system. It is a
propellant-based and a propellant-less system which create thrust by the
sun-driven ejection of a flux of particles of non-zero rest mass due to the
desorption of coating and solar radiation pressure, while it performs as a
propellant-less conventional solar sail after the thermal desorption ends.

In this paper we consider the dynamics of a deployed torus-shaped sail,
which can be unfurled using an inflatable torus-shaped rim structure. The
torus-shaped sail consists of a thin reflective membrane attached to an
inflatable torus-shaped rim. The sail's deployment from its stowed
configuration is initiated by introducing inflation pressure into the
toroidal rim with a round flat membrane coated by special heat-sensitive
materials that undergo the transition from the solid state phase into the
gas phase \cite{Benford}. The membrane is kept open and flat by the
distribution of the dynamic pressure of the gas in the toroidal rim. The
deployment and acceleration of the solar sail could be split into three
steps: in the first step, at a particular heliocentric distance the
torus-shaped solar sail is deployed by the gas inflated into the toroidal
structure and the membrane, which is coated with materials that undergo
thermal desorption (TD) at a specific temperature, is extended to a final
flat shape; in the second step, the membrane coat undergoes TD as a result
of heating by solar radiation and the inflation deployed torus-shaped solar
sail is accelerated via TD of coating \cite{1} as well as by solar radiation
pressure; in the third step, when the TD process ends, the sail utilizes
thrust only from the Sun and escapes the Solar System through the
conventional acceleration due to solar radiation pressure (SRP). We
determine the required structural strength of the inflatable torus to
support the flat surface of a circular membrane of a solar sail and study
the deflection of the flat membrane due to accelerations initiated by TD and
SRP using the equation of membrane. While the governing equation for a gas
in an inflated torus is assumed to be the equation for an ideal gas. Within
such approach we investigate the effects of both the enclosed gas pressure
and structure stiffness on the mode shapes of the inflated torus and the
stability of the torus-shaped solar sail. It is demonstrated that the effect
of the enclosed gas must be considered in the dynamic analysis of the
inflatable torus with the membrane.

The paper is organized in the following way. In Sec. 2 the dynamics of a
sail with thermal desorption of coating is considered. The solar sail
configuration is described in Sec. 3. In particular, it is considered outer
edge of a sail - a toroidal rim, and characteristics of a toroidal shell. In
the next subsections we analyze the sail membrane deflection under the solar
radiation pressure and the pressure induced by the thermal desorption of the
coating material, as well as vibrations of the membrane is addressed, while
in Subsec. 3.3 is analyzed the pressure of the gas which fills the toroidal
rim required to unfurl the sail and provide a desired tensile strength. The
effect of the electrostatic pressure for deployment of the sail is discussed
in Subsec. 3.4. In Sec. 4 are given results of calculations and discussion.
Conclusions follow in Sec. 5.

\section{Dynamics of a sail with thermal desorption of coating}

Thermal desorption produces thrust via releasing atoms at high speed from
the surface of the coated sail and acts like a jet plate. This release is
assumed to occur normal to the coating surface.
Therefore, the coating mass varies and at any instant $t$ it can be
presented as $M_{c}(t)\equiv M_{c}=M_{0}-m_{0}t$, where $M_{0}$ is the total
coating mass and $\frac{dM_{c}(t)}{dt}=-m_{0}$ is the rate of desorption. It
is obvious that the total time of the acceleration due to the desorption is
equal to the time of the desorption, and thus $t_{D}=\frac{M_{0}}{m_{0}}$.
The coated unfolded sail is a moving object with variable mass due to
coating material loss during the desorption period. At the end of the
desorption process $t=t_{D},$ one has $M_{c}(t_{D})=0$, namely the coating
mass is completely desorbed. The fact of the variation of mass during the
desorption cannot be neglected and should be taken into consideration.
Moreover, for the stability of the sail, it is important to have a symmetric
coating mass distribution on the surface of the sail. The total
instantaneous mass of the torus-shaped solar sail with the area $A$ and
areal mass of the membrane $\sigma $ is $M(t)=\sigma A+M_{t}+M_{P}+M_{c}$,
where $M_{t}$ and $M_{P}$ are the mass of the torus filed with the gas of
mass $M_{g}$ and mass of the payload, respectively. Since the total mass of
the coated sail varies in time, the force on the sail can be written as

\begin{equation}
F=\frac{d}{dt}\left( M\left( t\right) v\right) =\left( \sigma
A+M_{t}+M_{P}+M_{c}\right) \frac{dv}{dt}+\frac{dM_{c}}{dt}v.  \label{eq:1}
\end{equation}%
However, in our case the solar sail utilizes thrust from the TD of coating
and solar radiation pressure. The force due to the thermal desorption is $F=%
\frac{dM_{c}}{dt}v_{th},$ where $v_{th}$ the thermal speed of desorpted
atoms of the coating material. According to Maxwell's electromagnetic
theory, solar electromagnetic radiation carries the energy and linear
momentum and the radiation pressure exerted on a surface due to momentum
transport by photons is $P_{s}=\frac{k}{r^{2}}$, and produces the resulting
force $\frac{k}{r^{2}}A,$ where $A$ is the surface area facing the sun and $%
k=\frac{\eta L_{s}}{2\pi c}.$ In the latter expression $L_{s}=3.842\times
10^{26}$ W is the solar luminosity, $c$ is the speed of light and $0.5\leq
\eta \leq 1.$ The case $\eta =0.5$ corresponds to the total absorption of
photons by the solar sail and $\eta =1$ corresponds to total reflection.
Therefore, the resultant force produced by the desorption and solar
radiation pressure is

\begin{equation}
F=\frac{dM_{c}}{dt}v_{th}+\frac{k}{r^{2}}A.  \label{Force}
\end{equation}%
By comparing equations (\ref{eq:1}) and (\ref{Force}) one obtains a
differential equation for $v$:

\begin{equation}
\frac{dv}{dt}-G(t)v+G(t)\left( v_{th}-\frac{kA}{r^{2}m_{0}}\right) =0.
\label{RK2}
\end{equation}%
In Eq. (\ref{RK2}) we introduce the time-dependent coefficient $G(t)$
defined as

\begin{equation}
G(t)=\frac{m_{0}}{\left( \sigma A+M_{t}+M_{P}+M_{0}-m_{0}t\right) }
\label{CoefRK}
\end{equation}%
by considering that $M_{c}=M_{0}-m_{0}t$ and $\frac{dM_{c}}{dt}=-m_{0}.$

We solve the first order differential equation (\ref{RK2}) with
time-dependent coefficient $G(t)$ assuming that when the torus-shaped sail is
deployed at the perihelion of the heliocentric orbit, where the mechanism of
the thermal desorption is turned on at $t=0,$ the sail's velocity is $v_{p},$
\textit{i.e.} $v(0)=v_{p}$ is the sail's velocity at the perihelion of the
heliocentric orbit. The corresponding solution of Eq. (\ref{RK2}) is

\begin{equation}
v(t)=G(t)\left[ \frac{v_{p}(\sigma A+M_{t}+M_{P}+M_{0})}{m_{0}}-\left(
v_{th}-\frac{kA}{r^{2}m_{0}}\right) t\right] .  \label{v(t)}
\end{equation}%
In Eq. (\ref{v(t)}) $v_{th}$ is the thermal speed of the desorbed atoms,
which depends on the temperature $T$\ of the membrane and the mass $m$ of
the atoms and is defined by the Maxwell speed distribution as

\begin{equation}
v_{th}=\sqrt{\frac{8k_{B}T}{\pi m}},  \label{Vth}
\end{equation}%
where $k_{B}=1.38\times 10^{-23}$ J$\cdot $K$^{-1}$ is the Boltzmann
constant. Therefore, the velocity of the sail is determined by the initial
velocity of the sail at the perihelion $v_{p},$ and depends on the rate of
desorption $m_{0}$, the thermal speed $v_{th}$ of the desorbed atoms, as
well as on the solar radiation pressure. By the end of the acceleration due
to desorption at $t_{D}=\frac{M_{0}}{m_{0}}$ the sail will gain the maximum
velocity

\begin{equation}
v_{\max }=v_{p}+\left( v_{p}-v_{th}+\frac{kA}{r^{2}m_{0}}\right) \frac{M_{0}%
}{\sigma A+M_{t}+M_{P}}.  \label{RKReal_Vmax}
\end{equation}%
Eq. (\ref{RKReal_Vmax}) shows that the maximal velocity of the sail is
determined by the initial velocity of the sail $v_{p}$ at the perihelion of
the heliocentric orbit, and ratio of coating mass $M_{0}$ to mass of the
sailcraft $\sigma A+M_{t}+M_{P}$ excluding the coating mass. With this
velocity, the sail continues to accelerate due to the solar radiation at a
lower rate but for a longer time interval. The corresponding description of
the sail acceleration due to the solar radiation is well known and can be
omitted here.

For the sake of simplicity in calculations of the maximum velocity using Eq.
(\ref{RK2}), one can neglect the contribution of the term $\frac{kA}{%
r^{2}m_{0}}$ related to the force due to solar radiation during the
desorption phase because the acceleration due to the thermal desorption is
considered to be a rapid enough process to allow this simplification. Of
course, acceleration due to solar radiation pressure increases the sail's
velocity when thermal desorption is occurring over close approach, but its
contribution is not significant because the desorption phase is so short,
only a few thousand seconds.
The corresponding analytical expressions for the velocity of the sail and
the reasonability of this approximation is discussed in details in Refs.
\cite{AnconaKezerashviliArxiv,AnconaKezerashvili2, AnconaKezerashvili3}.
Also one should mention that the high reflectivity you want in a good solar
sail is diametrically opposed to the high absorptivity you want in a coating
material that's going to evaporate off.

\section{Solar sail configuration}

Below we consider the main elements of the torus-shaped solar sail: the
toroidal rim, which deploys the solar sail, the sail's membrane, and the
required pressure of the gas within the toroidal shell.

\subsection{Outer edge of a sail: a toroidal rim}

\bigskip A toroidal rim is a key component and basic structural element
functioning to carry the load imposed by the inflation pressure as well as
take up the tensile forces created by the stretched membrane at the inner
edge and provides structural support to the sail. Here we consider a
structural static behavior of an inflated toroidal rim following Ref. \cite%
{Kraus1967}. We assume large aspect ratio, the thickness of the torus shell
is negligible compared to the radii of curvature, an inflated gas affects
moderately to out-plane loading condition with more peaks compared to that
of the in-plane loading condition and any change in length from deformation
is insignificant and can be neglected.

\begin{figure}[h]
\noindent
\begin{centering}
\includegraphics[width=6.8cm]{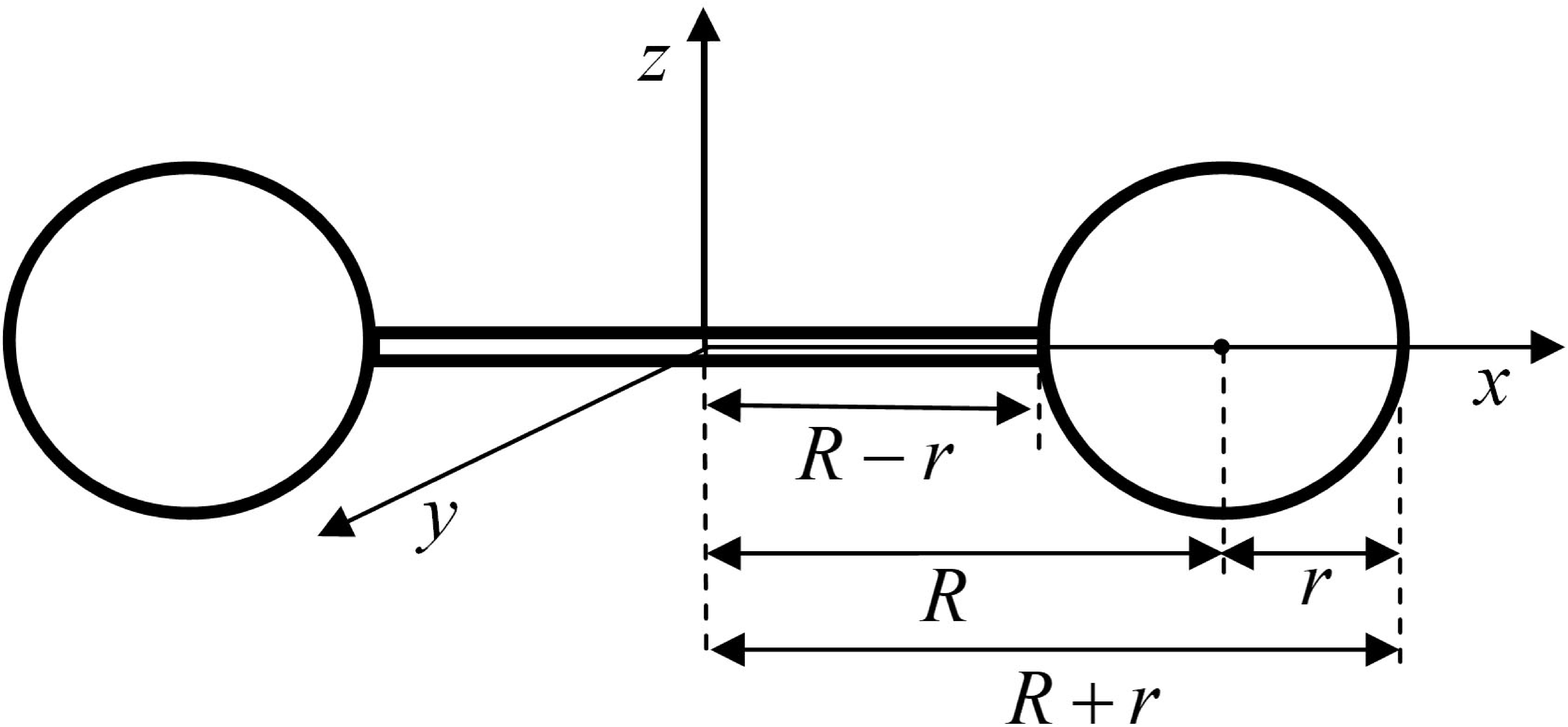}
\includegraphics[width=8.5cm]{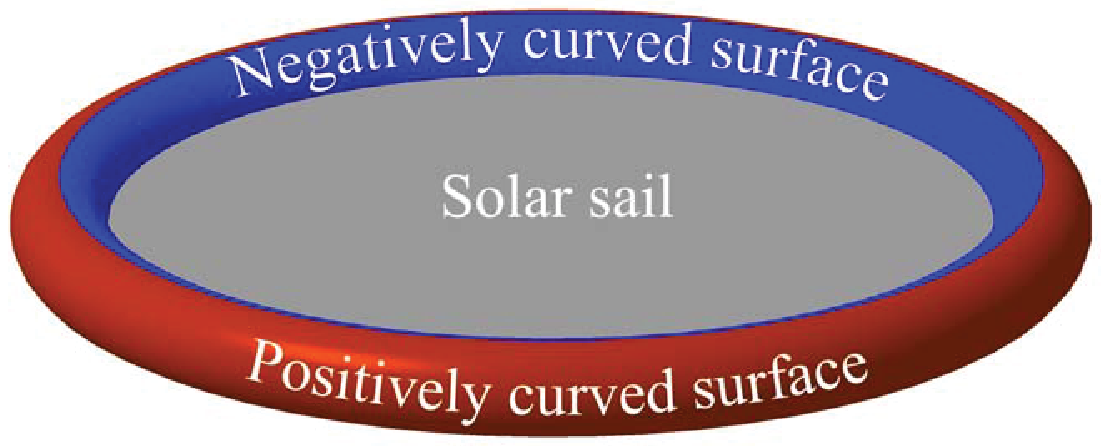}
\par\end{centering}
\caption{(Color online) Torus-shaped solar sail and its geometric
characteristics.}
\label{Torusfig}
\end{figure}

In the left panel in Fig. \ref{Torusfig} the torus with circular cross
section that is generated by revolving the circle $(x-R)^{2}+z^{2}=r^{2}$ of
radius $r$ in the $xz$ $-$ plane about the $z$ axis at the distance $R$ from
the center of the circle is shown. For a ring torus with a large aspect
ratio\ we have $R\gg r$. An implicit equation in Cartesian coordinates for a
torus azimuthally symmetric about the $z-$axis is

\begin{equation}
(R-\sqrt{x^{2}+y^{2}})^{2}+z^{2}=r^{2},  \label{torus1}
\end{equation}%
where $R$ is the distance from the center of the tube to the center of the
torus and $r$ is the radius of the tube.

\bigskip It is convenient to use toroidal and poloidal coordinates. The
polar coordinate $(r,\phi )$ at each cross section and circumferential
(toroidal) coordinate $\theta $ around the toroidal ring are chosen to
describe the geometric variation in the torus. The toroidal coordinate with
an origin at the center of the torus is used to describe those quantities
along the direction normal to the cross section. The three coordinate set $%
r,\phi ,\theta $ of curved coordinate system are orthogonal to each other.
Then the toroidal and poloidal coordinate system relates to standard
Cartesian coordinates with basic toroidal geometry are equated as

\begin{eqnarray}
x &=&(R+r\cos \phi )\cos \theta ,  \notag \\
y &=&(R+r\cos \phi )\sin \theta ,  \notag \\
z &=&r\sin \phi .  \label{Torcoordinate}
\end{eqnarray}

Our special interest is on the surface area of the torus. The total surface
area of the torus can be considered as a sum of the outer and inner
surfaces (right panel in Fig. \ref{Torusfig}). The torus have two curvatures in two principal direction causes
coupling between bending and stretching actions of a membrane. The outer
surface is a positively curved in a contrast of the inner surface, which is
negatively curved. The surface is positively curved, if a tangent plane
placed on any point on the outer surface, all the points on the surface of
the torus will be curving away from it on one side of the plane only,
whereas a negatively curved surface a tangent plane located on any point on
the inner surface, results of points on the torus that curve away on both
sides of the plane. Our interest is to find how much larger is the outer
surface than the inner one. Let us find surface area of both positively and
negatively curved surface areas. If the measure of the length is $%
s=r\phi ,$ thus $ds=rd\phi ,$ where $\phi \in \left[ \frac{\pi }{2},\frac{%
\pi }{2}\right] $. Therefore, the outer surface area of the torus is

\begin{equation}
A_{out}=\int_{-\frac{\pi }{2}}^{\frac{\pi }{2}}2\pi \left( R+x\right)
ds=\int_{-\frac{\pi }{2}}^{\frac{\pi }{2}}2\pi \left( R+r\cos \phi \right)
rd\phi =2\pi ^{2}Rr+4\pi r^{2}.  \label{outer}
\end{equation}%
For the area of negatively curved surface one obtains

\begin{equation}
A_{in}=\int_{-\frac{\pi }{2}}^{\frac{\pi }{2}}2\pi \left( R-x\right)
ds=\int_{-\frac{\pi }{2}}^{\frac{\pi }{2}}2\pi \left( R-r\cos \phi \right)
rd\phi =2\pi ^{2}Rr-4\pi r^{2}.  \label{inner}
\end{equation}%
Therefore, the positively curved surface is larger than the inner surface as
much as

\begin{equation}
\Delta A=8\pi r^{2},  \label{Differencearea}
\end{equation}%
while the surface area of the torus is

\begin{equation}
A=A_{out}+A_{in}=4\pi ^{2}rR.  \label{Sarea}
\end{equation}%
The interior volume of the torus can be computed from Pappus\'{\i}s centroid
theorem and is given by

\begin{equation}
V=2\pi ^{2}r^{2}R.  \label{Volume}
\end{equation}%
The volume of the filled gas is determined by the interior volume of the
torus and given by (\ref{Volume}).

Let us determine the mass of the sail. We assume that the surface of the
torus is made from a material with density $\rho _{t}$ and the density of
the material of the membrane, which has the area $A=4(R-r)^{2},$ is $\rho
_{m}$. If the thickness of the foil for the toroidal shell and membrane is
the same and equals $d,$ the mass of the sail is

\begin{equation}
M=\sigma A+M_{t}=4\pi ^{2}\rho _{t}rRd+\pi \rho _{m}(R-r)^{2}d+M_{g},
\label{Mass}
\end{equation}%
where $M_{t}$ is the mass of the toroidal shell filled with the gas of mass $%
M_{g}$.

\subsection{Sail membrane}

The sail surface is a circular membrane of radius $R-r$, which is considered
as a disk of radius $R-r$, and height $d$ centered at the origin, which will
represent the "still" drum head shape under the tensile strength produced by
toroidal rim. The boundary of the membrane is a circle of radius $R-r$
centered at the origin. Thus, the boundary is the inner part of the torus
and represents the rigid frame to which the membrane is attached. In the
equilibrium position the membrane is stretched and fixed along its entire
boundary by the inflated toroidal rim in the $xy-$plane. The tension per
unit length $q$ caused by stretching the membrane is the same at all points
and in all directions and does not change during the motion. The membrane is
perfectly flexible and offers no resistance to bending. At any time the
deflection of the membrane shape at a point is measured from this "still"
membrane shape will be denoted by $\zeta (\rho )$, which can take both
positive and negative values. Assuming that the shape in the loaded
condition is shallow enough to linearize the problem, the equilibrium
equation of the membrane in polar coordinates is \cite{Genta1999}

\begin{equation}
\frac{d^{2}\zeta (\rho )}{d\rho ^{2}}+\frac{1}{r}\frac{d\zeta (\rho )}{d\rho
}=-\frac{p}{q},  \label{membrane1}
\end{equation}%
where $\zeta (\rho )$ is the deflected configuration, $q$ is the in-plane
tensile force per unit length in the membrane and $p$ is the pressure acting
on the membrane. Eq. (\ref{membrane1}) has the analytical solution, yielding

\begin{equation}
\zeta (\rho )=\frac{p}{4q}\left[ (R-r)^{2}-\rho ^{2}\right] ,
\label{diflection}
\end{equation}%
where $R-r_{\text{ }}$is the outer radius of the sail.{}

During the thermal desorption acceleration, the membrane experiences both
the solar radiation pressure and the pressure due to desorption of the
coating material. The total pressure is

\begin{equation}
p=\frac{k}{r^{2}}+\frac{m_{0}v_{th}}{A},  \label{Totalpressure}
\end{equation}%
where the first and second terms in (\ref{Totalpressure}) correspond to the
pressure due to solar radiation and thermal desorption of coating material,
respectively. This pressure depends on the heliocentric distance, the rate
of the desorption process and the thermal speed of the desorbed atoms. At
the end of the desorption precess only the solar radiation pressure acts on
the membrane. The maximum deflection of the membrane is

\begin{equation}
\zeta _{\max }=\frac{1}{4q}\left( \frac{k}{r^{2}}+\frac{m_{0}v_{th}}{A}%
\right) .  \label{MaxDiflection}
\end{equation}%
From the other side when the desorption process ends the sail surface
suddenly experiences less pressure which will cause the vibration of the
surface of the sail. The mathematical equation that governs the vibration of
the membrane is the wave equation with zero boundary conditions. The wave
equation has been widely studied in the literature and is as follows:

\begin{equation}
\frac{\partial ^{2}u}{\partial t^{2}}=c_{s}^{2}\left( \frac{\partial ^{2}u}{%
\partial x^{2}}+\frac{\partial ^{2}u}{\partial y^{2}}\right) ,\text{ \ }%
c_{s}^{2}=\frac{q}{\sigma },  \label{waveeq}
\end{equation}%
where $u(x,y)=0$ on the boundary of the membrane, $\sigma $ is the mass per
area of the solar sail,\ and $c_{s}=\sqrt{\frac{q}{\sigma }}$ is the sound
speed. Eq. (\ref{waveeq}) is the two-dimensional wave equation, which is a
second order partial differential equation. Due to the circular geometry of
the membrane it is convenient to rewrite Eq. (\ref{waveeq}) using the
cylindrical coordinates

\begin{equation}
\frac{\partial ^{2}u}{\partial t^{2}}=c_{s}^{2}\left( \frac{\partial ^{2}u}{%
\partial \varrho ^{2}}+\frac{1}{\varrho }\frac{\partial u}{\partial \varrho }%
+\frac{1}{\varrho ^{2}}\frac{\partial u}{\partial \theta }\right) ,\text{ \
for }0\leq \varrho <R-r,\text{ and \ }0\leq \theta \leq 2\pi  \label{Radial}
\end{equation}%
with the boundary condition $u(\varrho ,\theta ,t)=0$ for $\varrho =R-r,$
which means that the membrane is fixed along the boundary circle of the
radius $\rho $\ in the $xy-$plane for all times $t\geq 0$. To determine
solutions $u(\varrho ,\theta ,t)$ that are radially symmetric, we solve Eq. (%
\ref{Radial}) following the three standard steps: i. using the method of
separation of variables, we first determine solutions as $u(\varrho ,\theta
,t)=U(\varrho ,\theta )V(t)$ and obtain two independent differential
equations for $U(\varrho ,\theta )$ and $V(t)$ functions; ii. from the
solutions of those ordinary differential equations we determine solution
(eigenfunctions) $U(\varrho ,\theta )$ which satisfy the boundary condition $%
U(\varrho ,\theta )=0$ for $\varrho =R-r$ and find the corresponding
eigenvalues. Then find the periodical solution $V(t)$ by solving the second
differential equation; iii. we compose these solutions and obtain the
radially symmetric solution $u(\varrho ,t)$ that satisfy the conditions $%
u(\varrho ,0)$ and $\frac{\partial u(\varrho ,t)}{\partial t}$ depend only
on $\rho .$ The corresponding solution is given in Appendix by Eq. (\ref%
{Total}).

\subsection{Torus filled with gas}

Now let us determine the required structural strength of the inflatable
torus to support the flat surface of a circular membrane of a solar sail. We
assume that at equilibrium the pressure of the gas in the torus obeys to the
isochoric process because it is confined in the given volume $V$\ of the
torus which only slightly changes shape due to the applied gas pressure and
the internal pressure remains normal to the surface. Seeking the simplicity
let us describe the gas by the equation for the ideal gas

\begin{equation}
PV=\frac{M_{g}}{\mu }R_{g}T,  \label{Klaiperon}
\end{equation}%
where $M_{g}$ is the required mass of the gas not considering diffusion
losses, $\mu $ and $T$ is the molecular weight and temperature of the gas,
and $R_{g}=8.31$ J$\cdot $K$^{-1}$mol$^{-1}$is a universal gas constant. For
the isochoric process one should maintain the pressure

\begin{equation}
P=\frac{M_{g}}{\mu }\frac{R_{g}T}{2\pi ^{2}r^{2}R}.  \label{pressure}
\end{equation}%
As it follows from Eq. (\ref{pressure}) the pressure of the confined gas is
defined by its mass and temperature, and characteristic size of the torus: $%
R $ and $r.$ Using Eqs. (\ref{Differencearea}) \ and (\ref{pressure}) one
can find the in-plane tensile force per unit length in the membrane produced
by the inflatable torus filled with gas

\begin{equation}
q=\frac{\Delta AP}{2\pi (R-r)}=\frac{2}{\pi ^{2}}\frac{M_{g}}{\mu }\frac{%
R_{g}T}{(R-r)R}.  \label{TencsilForce}
\end{equation}%
The gas confined in the torus has a high temperature and therefore it can
diffuse through the wall of the toroidal shell. Let us consider atomic
diffusion - the process whereby the random thermally-activated gas atoms
result in net transport through the toroidal foil shell. In case of hydrogen
gas, hydrogen molecules, atoms and ions inside the torus can diffuse through
the toroidal shell wall and escape, resulting in the sail slowly deflating.
There is a concentration gradient in the toroidal shell wall, because the
torus is filled initially with hydrogen, and there is no hydrogen on the
outside. The equation for a flux of hydrogen through the toroidal shell foil
is

\begin{equation}
\mathbf{j}=-D\left( \nabla C+C\frac{Q}{R_{g}T^{2}}\nabla T\right) ,
\label{Diffusion}
\end{equation}%
where $Q$ is the heat transport, while $D$ is the diffusivity, and $C$ is
the concentration which are measured in mol/m$^{3}$ and m$^{2}$/s,
respectively. Below we consider hydrogen as a gas and beryllium as a
material of the sail membrane and toroidal shell. In this case Eq. (\ref%
{Diffusion}) contains two hydrogen-beryllium interaction parameters, the
diffusivity $D$, which describes the hydrogen transport in a concentration
gradient and the heat transport $Q$, which describes the hydrogen transport
in a temperature gradient. Because the thickness of the toroidal shell wall\
is about a few tens of nanometers, the temperature gradient is negligibly
small and in Eq. (\ref{Diffusion}) to a very good approximation we can
consider that $\nabla T=0$. Therefore, the rate of transport of hydrogen is
governed by the diffusivity and the concentration gradient and a diffusion
flux can be determined by the Fick's first law \cite{Fick1} as

\begin{equation}
\mathbf{j}=-D\nabla C.  \label{Fick}
\end{equation}%
The diffusion coefficient $D$ of the hydrogen through the material depends
on the velocity of the diffusing particles, which in turn depends on the
temperature, and the size of the particles and is defined by an Arrhenius
equation $D=D_{0}\exp \left( -\frac{E_{D}}{k_{B}T}\right) ,$ where $D_{0}$
is diffusion constant, $E_{D}$ is the activation energy for the diffusion in
electronvolts (eV) and $k_{B}=1.38\times 10^{-23}$J$\cdot $K$%
^{-1}=8.62\times 10^{-5}$ eV/K is the Boltzmann's constant. The hydrogen
fill gas is in thermal equilibrium with the beryllium toroidal shell wall.
Therefore, the gas temperature is equal to the temperature of the beryllium
foil. In the best case scenario this temperature should be less than the
beryllium melting temperature 1551 K.

For simplicity let us consider the diffusion flux perpendicular to the wall.
Also the concentration of the hydrogen atoms deceases uniformly. Under these
assumptions and due to very small thickness $d$\ of the beryllium toroidal
shell wall as a good approximation we can replace the gradient of the
concentration by the concentration difference $C_{out}-C_{in}$ over
thickness $d$, where $C_{out}$ and $C_{in}$ are the concentration of
hydrogen inside and outside of the beryllium toroidal shell. However, the
outside concentration of the hydrogen $C_{out}=0$ and in this case the Fick
equation (\ref{Fick}) can be rewritten as follows

\begin{equation}
j=D_{0}e^{-\frac{E_{D}}{k_{B}T}}\frac{C_{in}}{d}.  \label{FickSimple}
\end{equation}%
Eq. (\ref{FickSimple}) shows the dependence of the hydrogen diffusion flux
on the temperature and thickness of the wall for the toroidal shell.

\bigskip The total flux rate of the hydrogen through the toroidal shell area
is%
\begin{equation}
\frac{dM_{g}}{dt}=\oint \mathbf{j\cdot n}dS,
\end{equation}%
where $\mathbf{n}$ is the normal to the toroidal surface $S$. Using Eqs. (%
\ref{Sarea})\ and (\ref{FickSimple}) we can estimated the total mass flux
rate as

\begin{equation}
\frac{dM_{g}}{dt}=4\pi ^{2}RrD_{0}e^{-\frac{E_{D}}{k_{B}T}}\frac{C_{in}}{d}.
\label{GasRate}
\end{equation}

\subsection{Electrostatic pressure}

Suppose that the toroidal shell surface is charged and the surface charge
density is $\sigma _{c}.$ The electric field just outside the charged torus
is $E$ and this field must be directed normal to the surface of the toroidal
shell. Any parallel component would be shorted out by surface currents.
Another way of saying this is that the surface of the toroidal shell, is an
equipotential surface. The electric field inside of the hollow-body toroidal
shell according to Gauss' law is zero, while at the surface of the toroidal
shell it is $E=\frac{\sigma _{c}}{\varepsilon _{0}}.$ In the presence of an
electric field, a surface charge on the toroidal shell will experience a
force per unit area which is an outward electrostatic pressure. The
electrostatic pressure can also be written as $P_{e}=\frac{\sigma _{c}}{2}E$%
, where $E$ is the electric field immediately above the surface of the
toroidal shell and is

\begin{equation}
P_{e}=\frac{\sigma _{c}^{2}}{2\varepsilon _{0}}.  \label{PressureElectrostat}
\end{equation}%
Note that the electrostatic pressure is equivalent to the energy density of
the electric field immediately outside the torus. Electrostatic pressure
acting on the toroidal surface produces the tensile force per unit length in
the membrane as follows from Eqs. (\ref{Differencearea}) \ and (\ref%
{PressureElectrostat}) is

\begin{equation}
q=\frac{\Delta AP_{e}}{2\pi (R-r)}=\frac{2}{\varepsilon _{0}}\frac{\sigma
_{c}^{2}r^{2}}{(R-r)}.  \label{TensilElectrost}
\end{equation}

The comparison of Eqs. (\ref{TencsilForce}) \ and (\ref{TensilElectrost})
shows that the charge density $\sigma _{c}\sim 2\times 10^{-3}$ C/m$^{2}$
will provide the same tensile force per unit length in the membrane as $%
0.3-0.5$ kg of hydrogen.

\bigskip The tensile strength of metals depends on the temperature. When
temperature increases the tensile strength of materials usually decreases
\cite{DepTemperature}. At room temperature the tensile strength of beryllium
is 370 MPa, while at temperature 1100 K its tensile strength decreases to
about 60 MPa \cite{MatalHandbook}. The pressure of the gas inside of the
toroidal shell when the temperature varies from 735 K ($r_{p}=0.3$ AU) to
1140 K ($r_{p}=0.1$ AU) is 0.12 MPa and 0.18 MPa, correspondingly. The same
electrostatic pressure corresponds to the surface charge distribution on the
toroidal shell $\sigma _{c}\sim 1.4\times 10^{-3}$ C/m$^{2}$ and $\sigma
_{c}\sim 1.8\times 10^{-3}$ C/m$^{2}$, respectively. Hence, in our case the
gas pressure or the electrostatic pressure does not exceeds the solar sail
material's tensile strength, and therefore the surface of the toroidal shell
will not fragment.

\section{Results and discussion}

The objective of the present work is to propose a deployment and study
dynamics of a sail, which can be unfurled using an inflatable torus-shaped
rim structure. We consider the following scenario. Using a conventional
spacecraft that carries the solar sails, the transfer occurs from Earth's
orbit to Jupiter's orbit. After that a Jupiter flyby leads to the
heliocentric orbit with the perihelion close to the Sun, where the
temperature corresponds the temperature of thermal desorption of the coating
material \cite{Ancona}. At this point the sail is deployed and thermal
desorption becomes active. During a short period, the sail is accelerated by
the thermal desorption and by solar radiation pressure. In our calculations
we consider the following parameter of the torus-shaped solar sail: the
radius of the toroidal rim and the torus tube of the sail \ is $R=10$ m and $%
r=0.2$ m, respectively. The reflected membrane area $A=301.72$ m$^{2},$
while the thickness of the membrane and the toroidal shell $d=40$ nm. For
this configuration the beryllium torus-shaped solar sail has the mass of
membrane $\sigma A=0.022$ kg, mass of the toroidal rim $4\pi ^{2}\rho
rRd=0.006$ kg, mass of the coating material $M_{0}=1.5$ kg with the rate of
the desorption $m_{0}=1\frac{\text{g}}{\text{s}},$ and we vary the molecular
hydrogen fill gas from 0.2 to 0.5 kg. Depending on the perihelion approach
the considered temperature range is $735-1140$ K.


\begin{figure}[h]
\centering
\includegraphics[width=0.49\columnwidth]{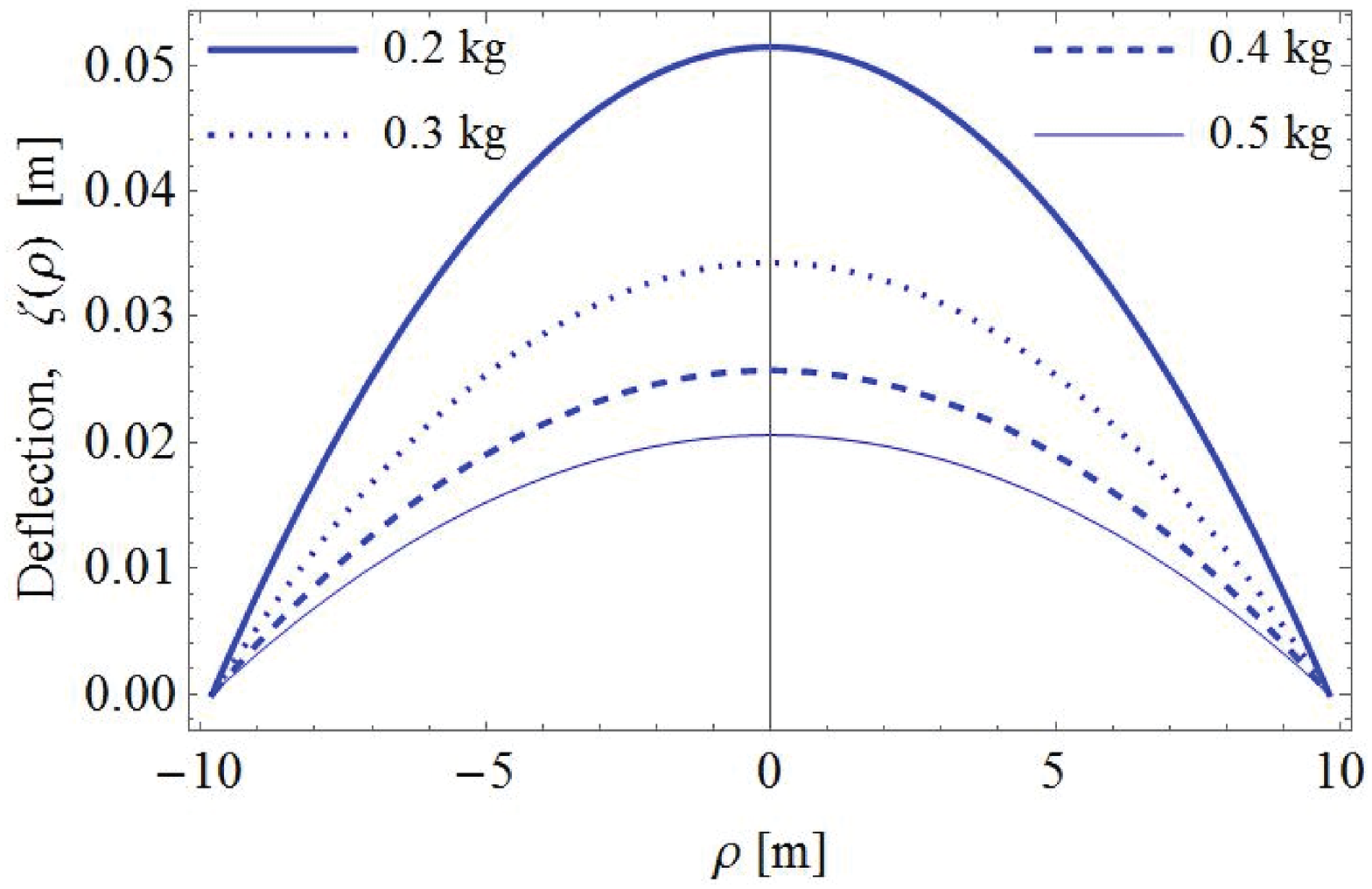} %
\includegraphics[width=0.49\columnwidth]{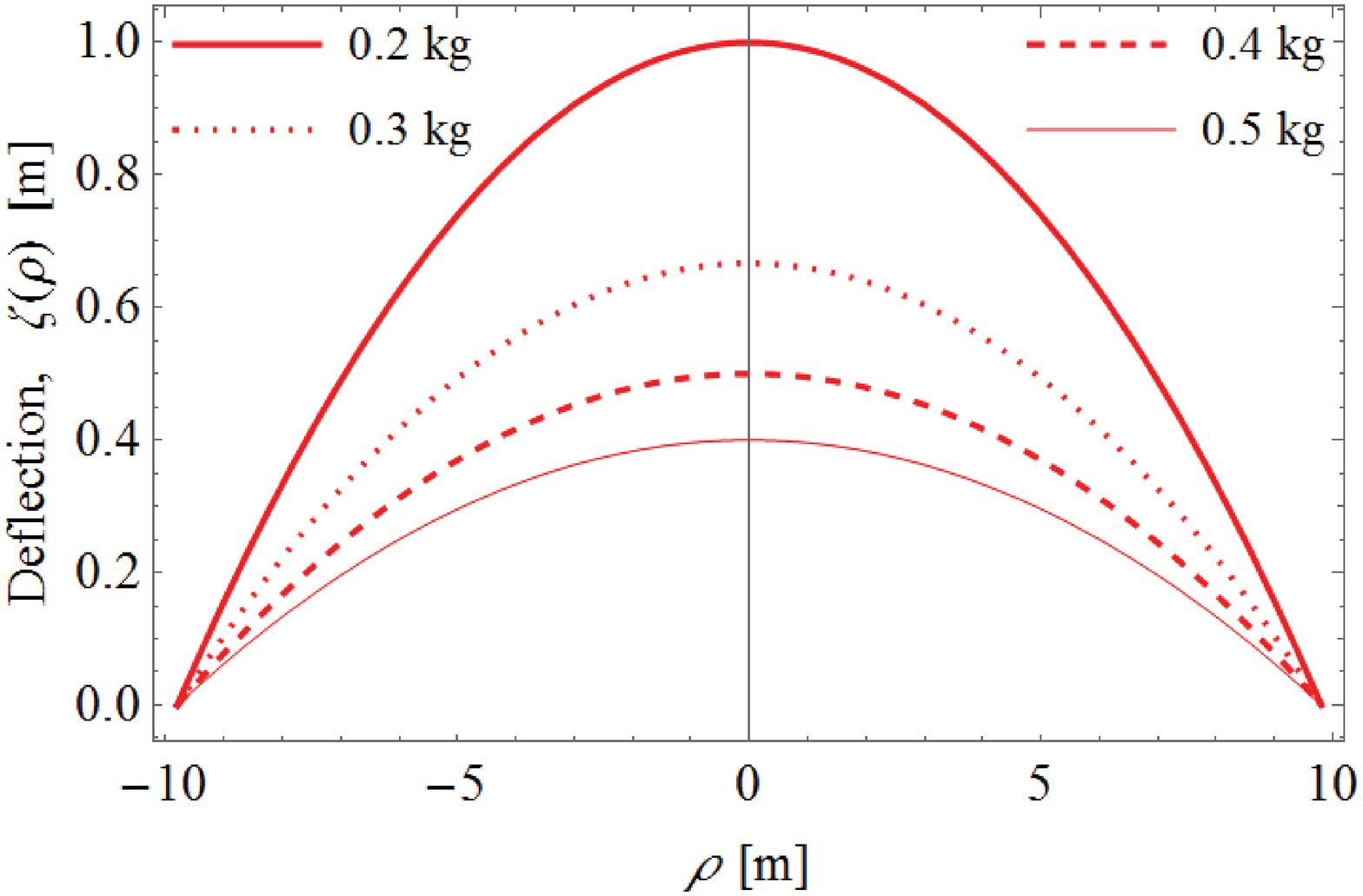}
\caption{Deflection of the membrane due to solar radiation (left panel) and
thermal desorption (right panel).}
\label{fig4}
\end{figure}

\begin{figure}[b]
\noindent
\begin{centering}
\includegraphics[width=6.0cm]{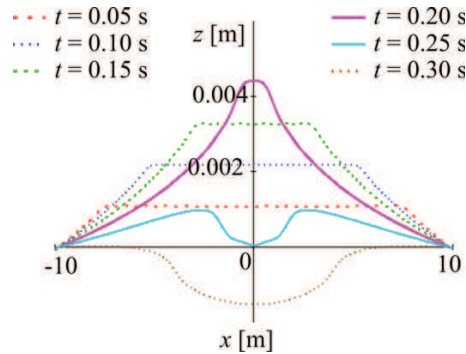}
\par\end{centering}
\caption{(Color online) Dynamics of the vibrating membrane.}
\label{2Dvibr}
\end{figure}
In Fig. 2 the results of the calculation of the deflection of the membrane
due to the solar radiation pressure and both the thermal desorption of
coating and solar radiation are shown. We perform calculations for different
masses of the molecular hydrogen fill gas. The different mass of the
confined gas lead to different pressure, which in turn changes the tensile
strength applied to the membrane. The increase of the pressure leads to the
increase of the tensile strength and, therefore, decreases the deflection of
the membrane. Under the solar radiation the membrane experiences only a few
cm deflection and maximum deflection from the flat position is varies from
0.08\% to 0.3\%. The deflection due to the both the thermal desorption and
solar radiation is significantly larger. However, the maximum deflections
for the different pressure of hydrogen gas is within a few percents only: $%
1.8\%-5\%.$ In both cases the maximum deflection of the membrane inversely
proportional to the mass of hydrogen fill gas.

We simulate the vibrations of the membrane under the following initial
conditions: at the initial time, the membrane has the initial deviations $%
\zeta _{\max }$ and all its points have an initial velocity of $1$ m/s. In
modeling we consider the sum up the 20 terms of the series (\ref{Total}),
which is equivalent to the sum of 20 tones of membrane oscillations. Figures
3 and 4 show the simulation results. Figure 3 shows the shape of the
deformed membrane surface for different instant of time. Figure 4 shows the
process of membrane oscillation in 3D space.

\begin{figure}[h]
\noindent
\begin{centering}
\includegraphics[width=15cm]{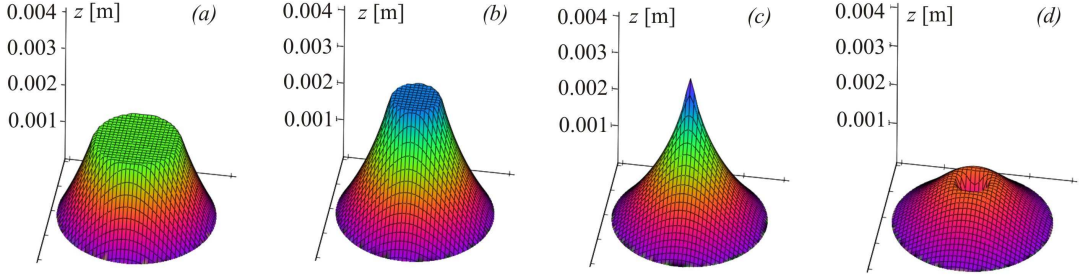}
\par\end{centering}
\caption{(Color online) 3D dynamics of the vibrating membrane. (a) at $t=0.1$
s; (b) $t=0.15$ s; (c) $t=0.20$ s; (d) $t=0.25$ s.}
\label{3DOscilat}
\end{figure}

\begin{figure}[b]
\noindent
\begin{centering}
\includegraphics[width=8.0cm]{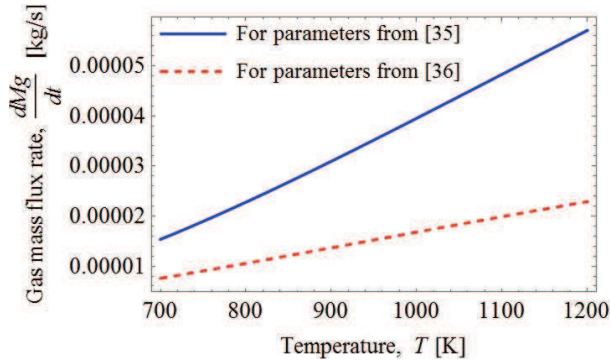}
\par\end{centering}
\caption{(Color online) Temperature dependence of hydrogen flux rate through
the toroidal beryllium shell. Results of calculation are given for the
different diffusion activation energy $E_{D}$ and diffusion constant $D_{0}$%
. }
\label{TemDepend}
\end{figure}

\bigskip Dynamics of an inflated torus is beyond the scope of present study
and the vibrational behavior of inflated tori has been extensively discussed
in the literature during the past six decades. Let us mentioned that
analysis on inflatable torus has been done in many publications and a
literature review \cite{Ruggiero} describes analytical as well as
experimental works on the dynamic response of inflated torus. Most recent
finite element analysis and effect of internal gas on modal analysis of an
inflatable torus are given in Refs. \cite{AdSpRes, China}

Let us consider the density of filled hydrogen gas. For the mass of gas $%
0.2-0.5$ kg the density varies from $2.53\times 10^{-5}$ kg/m$^{3}$ to $%
6.33\times 10^{-5}$ kg/m$^{3}.$ Thus, we are justifying that the hydrogen
gas confined by the toroidal shell can be considered as an ideal gas. Using
Eq. (\ref{GasRate}) we can determine the dependence the total flux rate of
the hydrogen on the temperature of the material. Fig. 5 presents the results
of our calculations for the dependence of the flux rate of the hydrogen on
the temperature for the different values of the diffusion activation energy $%
E_{D}$ and diffusion constant $D_{0}$. The experimental value of the
activation energy $E_{D}$ and diffusion constant $D_{0}$ for the diffusion
of beryllium depends on the purity of the beryllium material used in the
experiments and varies from 0.16 eV to 0.19 eV and from $9\times 10^{-12}$ m$%
^{2}$/s to $3\times 10^{-11}$ m$^{2}$/s, respectively \cite%
{Jones1967,Tazhibaeva1994,Causey}. A general overview of these results shows
that the diffused mass flux rate strongly depends on the temperature. When
the temperature increases from 700 K to 1200 K, the flux rate increases by a
factor of 5 and 2.5, when in calculations are used the diffusion activation
energy and diffusion constant from Refs. \cite{Jones1967} and \cite%
{Tazhibaeva1994}, respectively. Hence, we can conclude that the results also
demonstrated the strong dependence of the mass flux rate on the values of
the diffusion activation energy and diffusion constant. The acceleration
time of the torus-shaped solar sail nearby the Sun due to the thermal
desorption is about 1500 s for the mass of the coating material 1.5 kg and
desorption rate 1 g/s. Using these values we can estimate the performance
time for the solar sail. From Fig. 5 it is easy to see that\ at the end of
thermal desorption process only 85 g, which is about $28.5\%,$ of the
hydrogen filled gas will diffuse through the wall of the beryllium toroidal
shell of the sail for the temperature regime up to 1200 K, when we used in
our calculations the diffusion activation energy and diffusion constant from
Ref. \cite{Jones1967}. While only about 34 g ($11.4\%$) of the hydrogen
filled gas at the same temperature regime will diffuse through the wall of
the toroidal shell, when we used the diffusion activation energy and
diffusion constant from Ref. \cite{Tazhibaeva1994}. Therefore, we are on the
safe side because the hydrogen pressure which corresponds to 0.2 kg of the
gas still keeps the membrane flat with the maximum deflection of the
membrane about 5 cm as this follows from the left panel of Fig. 2. If one
will use 0.4 kg of hydrogen filled gas, the deflection of the membrane will
not exceed about 3 cm and will gradually decrease with the increase of
heliocentric distance due to the decrease of the solar radiation pressure.

In following preliminary analysis the planets and the Sun are considered
point-like. In the heliocentric reference frame, assuming that Earth's orbit
is almost circular, the sail has to be transferred to an inner orbit closer
to the Sun, in order to escape the Solar System. The transfer between these
two coplanar circular orbits is different and depends on the proposed
scenario. The results of calculations for a scenario for elliptical transfer
plus slingshot plus thermal desorption acceleration are presented in Table
1. For each perihelion of the heliocentric escape orbit the corresponding
speed at perihelion before desorption $v_{p}$, the maximum speed $v_{\text{%
max}}$ after desorption, the eccentricity of the hyperbolic heliocentric
orbit, temperature at the perihelion, cruise speed $v_{c}$ and distance $%
D_{y}$ covered by a solar sail per year are presented. Calculations are
performed for the coating mass $M_{c}$ = 1.5 kg, desorption rate $m_{0}$ = 1
g/s and mass of payload $M_{P}$ = 1.5 kg. \ The eccentricity of the
hyperbolic heliocentric orbit after the acceleration due to desorption ends
is estimated as $e=v_{\max }^{2}\frac{r_{p}}{\mu },$ where $\mu =1.327\times
10^{20}$ m$^{3}$s$^{-2}$ is the Sun's gravitational parameter.
\begin{table}[t]
\caption{Scenario for elliptical transfer plus slingshot plus thermal
desorption acceleration. For each perihelion of the heliocentric escape
orbit are presented the corresponding speed at perihelion before desorption $%
v_{p}$, the maximum speed $v_{\text{max}}$ after desorption, the
eccentricity of the hyperbolic heliocentric orbit, temperature at the
perihelion, cruise speed $v_{c}$ with out solar radiation pressure, $v_{sc}$
cruise speed including the solar radiation pressure, and distance $D_{y}$
covered by a solar sail per year without TD and $D_{s_{y}}$ is a distance
covered by a solar sail per year with TD. Calculations are performed for the
coating mass $M_{c}$ = 1.5 kg, desorption rate $m_{0}$ = 1 g/s and mass of
payload $M_{P}$ = 1.5 kg.}
\begin{center}
\begin{tabular}{ccccc}
\hline
\multicolumn{5}{c}{Characteristics of the mission} \\ \hline
$r_{p}$ & AU & 0.3 & 0.2 & 0.1 \\
$v_{p}$ & km/s & 73.24 & 91.23 & 129.11 \\
$T$ & K & 735 & 864 & 1140 \\
$v_{\max }$ & km/s & 133.23 & 166.20 & 235.35 \\
$e$ & -- & 1.01 & 2.12 & 5.26 \\
$v_{c}$ & km/s & 110.33 & 138.2 & 194.09 \\
$v_{sc}$ & km/s & 120.29 & 146.90 & 202.49 \\
$D_{y}$ & AU/year & 23.21 & 29.08 & 41.05 \\
$D_{s_{y}}$ & AU/year & 25.31 & 30.91 & 42.52 \\ \hline
\end{tabular}%
\end{center}
\end{table}

\section{Conclusions}

The advantage of inflation deployed torus-shaped solar sail accelerated via
thermal desorption is clearly evident. The present study reveals that the
inflation deployed torus-shaped solar sail accelerated via thermal
desorption of coating results in high post-perihelion heliocentric solar
sail velocities. With the speed 20$-$40 AU/year, post-perihelion travel
times to the vicinity of Kuiper Belt Objects\ (KBO) will be less than $1-3$
years, while the the Sun's gravity focus at 547 AU can be reached in $13-25$
years. We present the calculations for the Jupiter slingshot scenario but
this would not make the Jupiter flyby a critical piece of the concept.
Getting to perihelion can be accomplished a number of ways. The suggested
configuration of the torus-shaped solar sail fits the cube-scale size
configurations. Recent research reveals that much smaller sails could be
incorporated with highly miniaturized chip-scale spacecraft. It is quite
possible that a single dedicated interplanetary \textquotedblleft
bus\textquotedblright\ could deploy many cube-scale sails at perihelion.
Sequential deployment of a fleet of solar sails could be timed to allow
exploration of many small KBOs from a single launch. The natural
continuation of this work can be extended in the following directions: i.
detailed research on materials for thermal desorption at temperature
suitable for solar sailing; ii. consideration of the Sun as an extended
source of radiation; iii. study the influence of solar sail surface
oscillations on the motion of a spacecraft performing an interplanetary
flight.

\appendix

\section{Vibration of the membrane}

\label{apa}

Let us consider axisymmetric vibrations of the membrane. This means that the
deviations of any point of the membrane depends only on the distance to the
center of the membrane. Thus, we determine solutions $u(\varrho ,t)$ that
are radially symmetric. Based on this assumption and following Ref. \cite%
{Starinova}, there is no angle dependence and Eq. (\ref{Radial}) becomes
simpler and takes the following form:

\begin{eqnarray}
\text{ \ \ \ \ \ \ \ \ }\frac{\partial ^{2}u}{\partial t^{2}}
&=&c_{s}^{2}\left( \frac{\partial ^{2}u}{\partial \varrho ^{2}}+\frac{1}{%
\varrho }\frac{\partial u}{\partial \varrho }\right) ,\text{ \ }
\label{Noangle} \\
u(R-r,t) &=&0\text{ and for all \ }t\geq 0,\text{ }  \label{A2} \\
u(R-r,0) &=&f(\varrho ),\text{ }\frac{\partial }{\partial t}%
u(R-r,0)=F(\varrho )  \label{Condition1}
\end{eqnarray}%
The conditions (\ref{A2})-(\ref{Condition1}) in Eq. (\ref{Noangle}) means
that the membrane is fixed along the boundary circle $\varrho =R-r$ and the
initial deflection $f(\varrho )$ and the initial velocity $F(\varrho )$
depend only on $\varrho $, not on $\theta ,$ so that we can expect radially
symmetric solutions $u(\varrho ,t)$. The eigenfunctions $u_{k}(\varrho ,t)$,
obtained by solving equation (\ref{Noangle}), have the form

\begin{equation}
u_{k}(\varrho ,t)=\left[ a_{k}\cos (\lambda _{k}ct)+b_{k}\sin (\lambda
_{k}ct)\right] J_{0}(\lambda _{k}\varrho ),  \label{uk(rt)}
\end{equation}%
where $J_{0}(\lambda _{k}\varrho )$ are Bessel functions of zero order, $%
\lambda _{k}=\frac{\mu _{k}}{R-r}$ eigenvalues {}{}of the problem and $\mu
_{k}$ are the roots of the zero-order Bessel function: $J_{0}\left( \lambda
_{k}(R-r)\right) =0.$ The vibration of the membrane corresponding to $%
u_{k}(\varrho ,t)$ is called the $k$th normal mode. One can obtain the
solution of Eq. (\ref{Noangle}) that satisfies the initial conditions by
considering the series

\begin{equation}
u(\varrho ,t)=\sum\limits_{k=1}^{n}\left[ a_{k}\cos (\lambda
_{k}ct)+b_{k}\sin (\lambda _{k}ct)\right] J_{0}(\lambda _{k}\varrho ),
\label{Total}
\end{equation}%
where $a_{k}$ and $b_{k}$ must be the Fourier--Bessel series

\begin{eqnarray}
a_{k} &=&\frac{2}{J_{1}(\mu _{k})^{2}}\int_{0}^{1}xJ_{0}(\lambda _{k}x)f%
\left[ \left( R-r\right) x\right] dx, \\
b_{k} &=&\frac{2\left( R-r\right) }{c^{2}\mu _{k}J_{1}(\mu _{k})^{2}}%
\int_{0}^{1}xJ_{0}(\lambda _{k}x)F\left[ \left( R-r\right) x\right] dx,
\end{eqnarray}%
where $x=\frac{\varrho }{R-r}$. The latter expressions are obtained from the
condition for the fulfillment of boundary conditions.

\end{document}